\documentclass[a4paper,aps,pra,twocolumn,superscriptaddress,showpacs]{revtex4}
\usepackage{graphicx}
\usepackage{amsmath}
\usepackage{amsfonts}
\usepackage{color}
\usepackage{bm}
\usepackage{pdfcomment}
\usepackage{pgfplots}
\usepackage{subfigure}

\usepackage{placeins}

\graphicspath{{newfigs/}}
\graphicspath{{PRL_figs/}}

\newcommand{\comment}[1]{}

\newcommand{\beq}{\begin{equation}}
\newcommand{\eeq}{\end{equation}}

\usepackage{pgfplots}
\usetikzlibrary{plotmarks}

\begin{document}


\title{Nano-friction in cavity quantum electrodynamics}


\author{T. Fogarty}
\affiliation{Theoretische Physik, Universit\"at des Saarlandes, D-66123 
Saarbr\"ucken, Germany}

\author{C. Cormick}
\affiliation{IFEG, CONICET and Universidad Nacional de C\'ordoba, Ciudad Universitaria,
X5016LAE, C\'ordoba, Argentina}

\author{H. Landa}
\affiliation{Univ.~Paris Sud, CNRS, LPTMS, UMR 8626, Orsay 91405, France}
\author{Vladimir M. Stojanovi{\'c}}
\affiliation{Department of Physics, Harvard University, Cambridge, MA 02138, USA}

\author{E. Demler}
\affiliation{Department of Physics, Harvard University, Cambridge, MA 02138, USA}

\author{Giovanna Morigi}
\affiliation{Theoretische Physik, Universit\"at des Saarlandes, D-66123 
Saarbr\"ucken, Germany}

\date{\today}

\begin{abstract}
The dynamics of cold trapped ions in a high-finesse resonator results from the interplay between the long-range Coulomb repulsion and the cavity-induced interactions. The latter are due to multiple scatterings of laser photons inside the cavity and become relevant when the laser pump is sufficiently strong to overcome photon decay. We study the stationary states of ions coupled with a mode of a standing-wave cavity as a function of the cavity and laser parameters, when the typical length scales of the two self-organizing processes, Coulomb crystallization and photon-mediated interactions, are incommensurate. The dynamics are frustrated and in specific limiting cases can be cast in terms of the Frenkel-Kontorova model, which reproduces features of friction in one dimension. We numerically recover the sliding and pinned phases. For strong cavity nonlinearities, they are in general separated by bistable regions where superlubric and stick-slip dynamics coexist. The cavity, moreover, acts as a thermal reservoir and can cool the chain vibrations to temperatures controlled by the cavity parameters and by the ions phase. These features are imprinted in the radiation emitted by the cavity, which is readily measurable in state-of-art setups of cavity quantum electrodynamics. 
\end{abstract}


\maketitle
\par

Cavity quantum electrodynamics (CQED) with cold atomic ensembles provides exciting settings in which to study the physics of long-range interacting systems \cite{Ritsch,Cataliotti}. The latter are found when the interparticle potential in $D$ dimensions exhibits a scaling with the distance $r$ slower than $1/r^D$  \cite{Campa}. This property is of relevance from the nuclear scale to astrophysical plasmas and leads to non-additivity of the energy, whose consequences are, amongst others, ensemble inequivalence and metastable states with diverging lifetimes \cite{Campa}. 

The dynamics of atoms in single-mode high-finesse resonators exhibits several analogies with well-known theoretical models of the statistical mechanics of long-range interacting systems \cite{Schuetz}. High-finesse cavities, in fact, trap photons for a sufficiently long time so that multiple scatterings can occur among the atoms inside the resonator. This  gives rise to an effective interatomic potential whose range can scale with the size of the system \cite{Rempe} and that can induce spontaneous atom ordering \cite{Black:2003,Baumann:2010} even in one dimension \cite{Domokos:2002}. Additionally, the system is intrinsically out-of-equilibrium since the resonator dissipates light, thus non-trivial phases are observed only in the presence of an external drive. Under these premises, photon shot noise gives rise to global retardation effects, which can effectively cool the atomic motion \cite{Ritsch,Domokos:2002,Schuetz}. 

\begin{figure}[h!]
\includegraphics[width=3in]{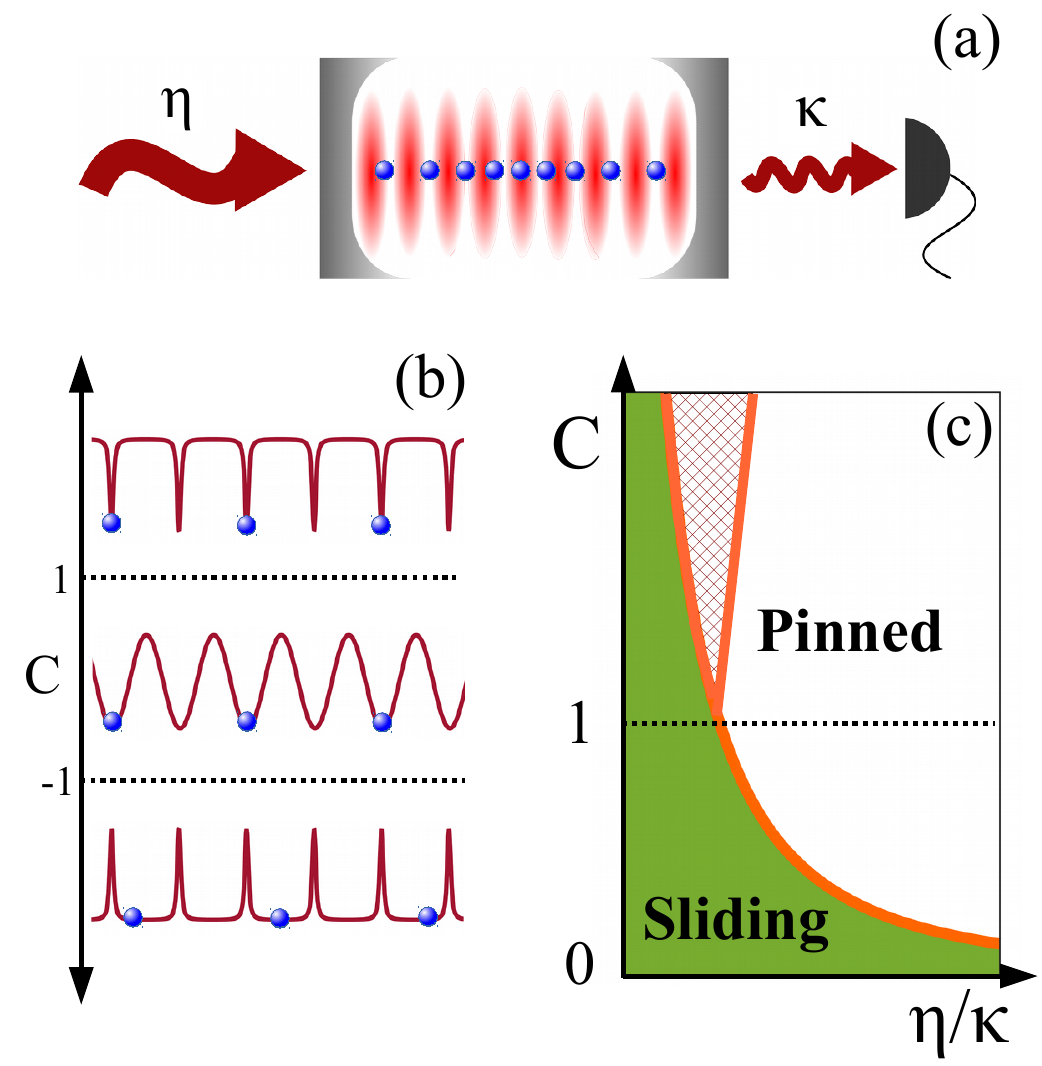}
\caption{(a) An array of cold ions in the optical lattice of a high-finesse cavity is an exotic realization of the 
Frenkel-Kontorova model. The cavity is pumped by a laser with amplitude $\eta$ and decays at rate $\kappa$, the intracavity photon 
number is determined by the ion density: this gives rise to a globally deformable potential which depends nonlinearly on the ions positions.
(b) The functional form of the cavity-induced potential for one particle: In the pinned phase, for $|C|<1$ the shape is 
approximately sinusoidal, for $|C|>1$ it becomes flat everywhere except in the vicinity of minima ($C>0$)
or maxima ($C<0$). (c) Sketch of the phase diagram for the stationary state as a function of $\eta$ and of the cavity nonlinearity $C$ (cooperativity). 
Typically for $|C|>1$ bistable phases can be observed (hatched region) signifying that superlubric or stick-slip dynamics are found depending on the variation of $\eta$ with time.}
\label{fig:1}
\end{figure}

The interparticle forces in general compete with these dynamics. When the interactions are short-ranged, at ultra-low temperatures their interplay with the cavity potential  can give rise to exotic phases, which tend to maximize photon scattering into the resonator and thus the strength of the long-range intracavity potential \cite{Ritsch,Larson2008,Baumann:2010,Gopala,Habibian:2013}. It is to a large extent unknown, however, how these dynamics are modified as the range of the competing interacting potential is increased. This question acquires further relevance in view of experimental setups trapping cold ions within high-finesse cavities \cite{Leroux,Cetina,Vuletic,Northup,Northup2,Yelin,Northup3}.

In this Letter we theoretically characterize the effect of cavity back-action in the presence of the competing Coulomb interaction between $N$ ions with the same charge $q$ and mass $m$. The ions are confined by an external trap inside a standing-wave resonator of wavelength $\lambda$, in the geometry illustrated in Fig. \ref{fig:1}(a), where their motion is assumed to be one-dimensional along the $x$-axis. This setup is expected to simulate the Frenkel-Kontorova (FK) model \cite{FK,Braun,Mukamel} which describes a chain of elastically-bound particles subjected to an external periodic potential (substrate) in one dimension \cite{Mata,Haeffner,Tosatti,Mandelli,Cetina,Vuletic} and reproduces the salient features of stick-slip motion between two surfaces. When the periodicity $\lambda/2$ of the cavity optical lattice is incommensurate with the characteristic interparticle distance $d$ of the ions (and the cavity nonlinearity is negligible), the ions ground state can be either sliding or pinned: In the sliding phase the forces giving rise to sticking can cancel, so that the minimal force for initiating sliding vanishes. Static friction becomes significant when ions are pinned by the cavity potential, in this case their distribution can still be incommensurate with the lattice periodicity, exhibiting defects (kinks). In the FK model the sliding and pinned phases are separated by the Aubry transition, whose control field is the relative amplitude of the periodic potential \cite{Aubry,Joos,Klafter}, and whose hallmark is the abrupt growth of the lowest phonon frequency (phonon gap) \cite{PeyrardAubry}. In a finite chain with free ends and an inversion-symmetric potential, this transition is characterized by symmetry breaking \cite{Mukamel,Klafter}. 

This behaviour is substantially modified when the periodic potential of the resonator mediates a long-range interaction among the atoms in the dispersive regime of CQED \cite{Kimble,Ritsch}.  Here, the coupling at strength $g$ to a cavity field mode, with spatial mode function $\cos(k x)$ and $k=2\pi/\lambda$, induces the conservative potential $\hbar U_0\hat n\cos^2(kx)$ for sufficiently large detuning $\Delta_0=\omega_c-\omega_{\rm el}$ between the frequencies of the cavity mode and ion dipolar transition. Then, the potential amplitude is proportional to the intracavity photon number operator $\hat n$. The strong-coupling regime of CQED is reached when the dynamical Stark shift per atom $U_0=g^2/\Delta_0$ is such that $N|U_0|/\kappa>1$, with $\kappa$ the cavity loss rate \cite{Kimble,Ritsch}. In this regime, the mean intracavity-photon number $\bar n=\langle \hat n\rangle$, and thus the depth of the cavity optical lattice, is a nonlinear function of the ions positions \cite{Cormick}: 
\beq
\label{nmean}
\bar n=|\eta|^2/(\kappa^2+\Delta_{\rm eff}(\{ x_j \})^2)\,,
\eeq
where $\eta$ is the amplitude of the driving field and 
\begin{equation}
\label{Delta_eff}
\Delta_{\rm eff}(\{ x_j \}) = \Delta_c -NU_0 B_N(\{ x_j \})\,.
\end{equation}
The detuning $\Delta_c=\omega_p-\omega_c$ of the pump from the cavity frequency is thus shifted by $NU_0B_N(\{ x_j \})$, where the function $B_N=\sum_j \cos^2(k x_j)/N$ depends on the ions positions $x_j$ along the cavity axis and is the so-called bunching parameter, as it measures their localization at the potential minima. This frequency shift is at the origin of the  nonlinear dependence of $\bar n$ on $\{x_j\}$ and gives rise to a deformation of the effective potential the ions experience, as shown in Fig. \ref{fig:1}(b), which can be expressed in terms of the effective mean-field potential
\begin{equation}
\label{V:cav}
V_{\rm cav}=-(\hbar|\eta|^2/\kappa) \arctan(\Delta_{\rm eff}(\{ x_j \})/\kappa)\,,
\end{equation}
and whose derivation is reported in Refs. \cite{Cormick,SI}. Its functional dependence is reminiscent of the nonlinearly deformable potential discussed in Refs. \cite{Peyrard,Pantic}. Potential $V_{\rm cav}$, however, also mediates a multi-body long-range interaction between the ions and thus acts as a globally deformable potential, since the potential depth depends on the global variable $B_N$.  It competes with the potential $V_{\rm ion}$ due to the Coulomb repulsion within the external harmonic trap, which orders the ions along the $x$-axis  \cite{Dubin_RMP,Morigi}, and whose axial component reads 
\begin{equation}
V_{\rm ion}=\frac{1}{2}\sum_{j=1}^Nm \omega^2x_i^2+\frac{q^2}{4\pi\epsilon_0}\sum_{j>i}\frac{1}{\vert x_j-x_i\vert}\,,
\end{equation}
with $\omega$ the trap frequency along $x$. In the absence of the cavity the equilibrium positions $\{x_i^{(0)}\}$ form a chain, the interparticle distance $d_j=x_{j+1}^{(0)}-x_j^{(0)}$ is inhomogeneous but almost uniform at the chain center \cite{Dubin97}, where it takes the minimal value $d$. Within the resonator the ions equilibrium positions $\{\bar x_j\}$ are the minima of the total potential $V=V_{\rm ion}+V_{\rm cav}$. We choose the trap frequency $\omega$ to ensure an incommensurate ratio between $\lambda$ and $d$, such that the dynamics are intrinsically frustrated. 

\begin{figure*}
\includegraphics[width=7.1in]{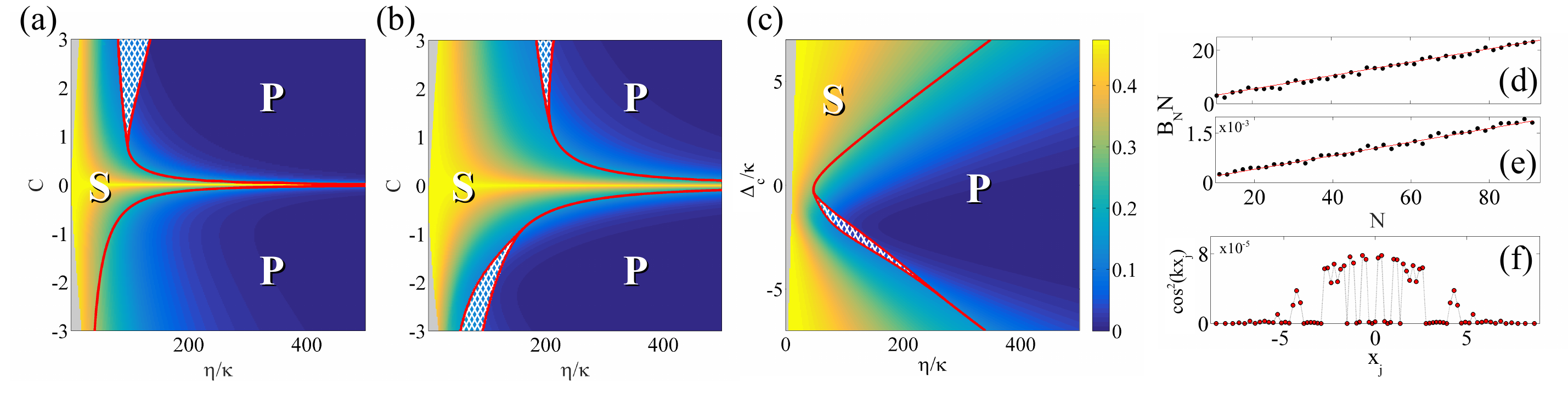}
\caption{Phase diagram as a function of $C$ and $\eta$ for (a) 
$\Delta_c=0$, (b) $\Delta_c=-2\kappa$. In (c) the phase diagram is plotted as a function of $\Delta_c$ and $\eta$
for cooperativity $C=-2$. Parameters $\eta$ and $\Delta_c$ are in units of $\kappa$. The solid red line indicates the 
symmetry breaking transition from the sliding (S) to the pinned (P) phase, the 
hatched white region is an area of bistability. The color code gives the value of $B_N$  for $C>0$, and of  $1-B_N$ when $C<0$. Subplots 
(d, e) display $B_NN$ vs. $N$ for $C=0.5$,  $\Delta_c=0$, and (d) $\eta=50\kappa$ and (e) $\eta=500\kappa$ 
respectively. (f) The individual contributions $\cos^2(kx_j)$ to $B_N$ for $N=81$ 
particles and deep in the pinned phase ($B_N=1.5\times 10^{-3}$) for $C=0.5$. The ions are $^{174}$Yb$^+$, 
the parameters are $|\Delta_0|=2\pi \times 12 \mbox{GHz}$, $\kappa=2\pi\times 0.2 \mbox{MHz}
$, for $N=11$ the trap frequency is $\omega=2\pi \times 1.12 \mbox{MHz}$, while $g$ is varied in order to 
sweep over different values of $C$. The cavity wavelength is $\lambda=369\mbox{nm}$ and the ratio 
$2d/\lambda=7.3507$. The centre of the harmonic trap corresponds with a maximum of $V_{\rm cav}$,  
ensuring a symmetry breaking transition. The 
grey area in (a-c) indicates where the mean phonon number $\bar n<1$. Outside of this region 
our semiclassical analysis is reliable. 
}
\label{Fig:2}
\end{figure*}

The strength of the cavity-mediated interactions is controlled by the cooperativity $C=NU_0/\kappa$, which scales the strength of the nonlinear shift in Eq. \eqref{Delta_eff}. For $|C|\ll 1$ the mean photon number is independent of the ions positions. In this limit $V_{\rm cav}\approx V_0\sum_j \cos^2(k x_j)$ and the total potential $V$  can be mapped to the FK model \cite{Mata,Haeffner,Vuletic}. The sliding-to-pinned transition is then expected at a critical value of the potential amplitude $V_0^c\propto \eta^2 C$, and occurs at smaller values of $\eta$ with increasing $|C|$, as illustrated in Fig. \ref{fig:1}(c). The sign of $U_0$, and thus of $C$, determines the features of the pinned phase: For $C>0$ a pinned configuration minimizes $B_N$, since the minima of $V_{\rm cav}$ are at the nodes, while if $C<0$ it maximizes $B_N$. As $|C|$ increases, the cavity potential changes shape, as in Fig. \ref{fig:1}(b), and it strongly depends on the value of $B_N$ through the shift in Eq. \eqref{Delta_eff}: For a fixed detuning $\Delta_c$, the resonance $\Delta_{\rm eff}=0$ is fulfilled for certain values of $B_N$, and thus for specific phases. For $|\Delta_c|>\kappa$, the resonance  can directly separate the regime where the minima are either spikes or flat bottomed.

In order to evaluate the ions phase we define an appropriate thermodynamic limit: Since $C\propto N$, we scale $U_0\sim 1/N$  to ensure that the resonance is at the same value of $B_N$ \cite{Fernandez-Vidal}. To fix the ratio $d/\lambda$, the trap frequency is scaled as $\omega\sim \sqrt{\log(N)}/N$ \cite{Morigi}. The equilibrium positions of the ions in the total potential are numerically determined as a function of $\eta$, $C$, and $\Delta_c$, and their stability is checked by means of a linear stability analysis.  We characterize the stationary state by determining the corresponding bunching parameter and by classifying it in terms of sliding or pinned phase. For this purpose, we compare the minimum phonon frequency \cite{Mata,Haeffner} and an analog of the depinning force in the classical FK model \cite{Tosatti,Haeffner,Vuletic}, which is evaluated following the procedure discussed in Ref. \cite{Tosatti,SI}.  For $C>0$, for instance, 
the trap center ($x=0$) is at a maximum, so for an odd number of ions, the sliding to pinned transition can be observed through the sudden displacement of the central ion from the origin. For larger $C$ a finite force is required to restore the symmetry of the system. At this transition we verify that the phonon gap is zero, and in the pinned regime it increases monotonically. Analogous considerations apply for $C<0$, where the maxima of the potential are at the antinodes of the field (in this case the center of the trap is shifted by $\lambda/4$ to obtain a symmetry breaking). We check that the values of $\eta$ and $C$ found by the symmetry breaking transition are the same at which the phonon gap starts to increase \cite{SI}. 

The resulting phase diagram is shown in Figs. \ref{Fig:2}(a) and (b) for 11 ions and as a function of $\eta$ and $C$ for  $\Delta_c=0$ and $\Delta_c=-2\kappa$ respectively, using the parameters of Ref. \cite{Cetina}. It exhibits sliding (S), pinned (P), and bistable phases (hatched). The color code gives the corresponding value of $B_N$. We have checked that the diagram remains substantially unvaried as the number of ions is scaled up according to the thermodynamic limit, apart from the bistable phases at $C>0$, as we discuss below. We first consider the transition line, delimiting the S phase. This moves to smaller values of $\eta$ as $|C|$ increases. For $|C|< 1$, the back-action due to the resonator can be neglected and the line follows the expected behaviour for the FK limit, with $\eta=\eta_c\sim 1/\sqrt{|C|}$. Here, it separates the S from the P phase and exhibits the typical features of the Aubry transition. At larger values it changes functional dependence. Moreover, sliding and pinned phases can coexist for $|C|>1$ about the transition line, a typical feature of a first order transition.

The bistable areas  at $C>0$, however, are of different nature than the one for $C<0$. For $C>0$ they are due to finite-size effects. As $C$ increases, in fact, the effective cavity potential becomes flat except for the nodes, where it exhibits tight minima, see  Fig. \ref{fig:1}(b). Thus,  this potential supports a stable sliding phase which is symmetric about the center, where the ions in general experience a flat potential. As $N$ is varied, bistability is still observed but it qualitatively changes its features. The bistable region in Fig.~\ref{Fig:2}(b) for $C<0$ is instead due to the resonance $\Delta_{\rm eff}=0$ and remains unvaried as $N$ is scaled up according to our prescription. This resonance occurs for specific values of $B_N$, and thus for specific sets of $\{\bar x_j\}$. Analogous resonances have been reported in experiments with cold atoms in resonators \cite{Hemmerich,Stamper-Kurn,Esslinger} and in theoretical works on similar setups \cite{Larson2008,Cormick,Meystre}. In our case they indicate that either superlubric or stick-slip behaviors can be encountered depending on how the intensity $\eta$ is varied in time. 
Fig. \ref{Fig:2}(c) shows the phase diagram for $C=-2$ and as a function of $\eta$ and $\Delta_c$. For this parameter choice, the resonance $\Delta_{\rm eff}=0$ exists only for $\Delta_c<0$, with values varying between the minimum and the maximum value of  $-2 B_N \kappa$, and thus of the bunching parameter.

The bunching parameter is a particularly relevant quantity, since its value can be extracted from the intensity of the light at the cavity output.  In the sliding phase and for $N\gg 1$, the particles are positioned at every point (modulus $\lambda$) of the cavity potential and thus $\lim_{N\to\infty}B_N=0.5$. In a commensurate phase $B_N\to 0$ for $C>0$ ($B_N\to 1$ for $C<0$).  At small deviations from this limiting value, $B_N$ is a crude estimation of the kink density.
The plots in Figs.~\ref{Fig:2}(a) and (b) show that $B_N$ signals the transition from sliding to pinned when $|C|\lesssim0.5$, however this is not the case for larger $C$. So in general for $B_N\sim 0.5$ the phase is sliding, while for $B_N<0.05$ at $C>0$ ($1-B_N<0.05$ at $C<0$) the phase is tightly pinned.
The estimated kink number $N_k\sim  B_NN$ grows linearly with $N$, as shown in Fig.~\ref{Fig:2}(d)-(e): The slope decreases 
as $\eta$ and $C$ are increased, but never vanishes, thus showing that the pinned phase remains incommensurate. Nonetheless, deep in the pinned phase the ions at the chain edges, where the density is smaller, organize at  commensurate distances with $\lambda$, as depicted in Fig.~\ref{Fig:2}(f). This effect results from the choice of harmonic confinement and it shows that the edge ions enforce a boundary condition restricting the central ones from becoming truly commensurate. 

Further features of the cavity long range interaction manifest in the ions vibrations, as they give rise to fluctuations in the potential, which in turn affects the ions motion. We analyze them in the linear regime and we denote $\delta \hat a$ and $\delta\hat x_j$ as the quantum fluctuations of the cavity annihilation operator and of the ion positions about the mean values $\bar a=\sqrt{\bar n}$ and $\bar x_j$, respectively, where $\bar n$ is given in Eq. \eqref{nmean} and $\bar x_j$ are the minima of $V$. We decompose the ions' displacement $\delta\hat x_j$ in the normal modes $\hat q_n=(\hat b_n+\hat b_n^\dagger)/\sqrt{2}$ calculated at zero order in $\delta \hat a$, with $\hat b_n$ the bosonic operator annihilating a chain phonon at frequency $\omega_n$. Cavity and phonons are coupled by the linearized Heisenberg-Langevin equations  in the presence of noise due to cavity decay and to an external damping reservoir coupled with the motion \cite{Cormick,SI}
%
%
\begin{eqnarray}
\label{eq:a}
\!\!\!& \delta \dot{\hat a} = ({\rm i} \Delta_{\rm eff} - \kappa) \delta \hat a - {\rm i} \bar a \sum_n c_n (\hat b_n + \hat b_n^\dagger) + \sqrt{2\kappa}\, \hat a_{\rm in}\,, \\
\label{eq:b}
\!\!\!& \dot {\hat b}_n = - ({\rm i} \omega_n + \Gamma_n) \hat b_n - {\rm i} \bar a c_n (\delta \hat a + \delta \hat a^\dagger) + \sqrt{2\Gamma_n} \, \hat b_{{\rm in}, n}\,,
\end{eqnarray}
where $c_n$ denotes the cavity coupling with mode $n$ and $\Gamma_n$ is the mode's damping rate. The Langevin operators $\hat \zeta_{in}=\hat a_{\rm in},\hat b_{{\rm in}, n}$ have zero mean value and $\langle [\hat{\zeta}_{\rm in}(t'), \hat{\zeta}_{\rm in}^\dagger (t'')] \rangle = \delta (t'-t'')$. The solutions are stationary when the eigenvalues possess no positive real parts. For $\Gamma_n=0$ the stability is determined by the cavity parameters and is warranted when $\Delta_{\rm eff}<0$ (the stability diagrams for the plots in Fig.~\ref{Fig:2}(a,b) are in the supplementary material \cite{SI}). In this regime, retardation processes in photon scattering cool the chain to effective temperatures $T$ which depend on the detuning $\Delta_c$ and on the bunching parameter $B_N$. Thus, the ions stationary state determines the temperature at which the chain is cooled. In turn, for a given $\Delta_c$ disparate regions in Fig.~\ref{Fig:2}(a-c) are generally at different temperatures since $B_N$ and $C$ vary.  From Eq.~\eqref{Delta_eff} one sees that for $\Delta_c<0$ and $C>0$ the ions are always cooled. Cooling for $C<0$ is found by suitably modifying the cavity detuning. In order to estimate $T$ we cast the modes in the form of a covariance matrix: For the parameters of Ref. \cite{Vuletic} and $N=11$ ions we find that $T\sim 125\mu$K can be reached. In this limit the standard deviation from the equilibrium positions, $\langle \delta \hat x_j^2\rangle^{1/2}$, is smaller than $\lambda/2$ in the pinned phase, thus the classical equilibrium positions dictate the phases of the system. This condition can be achieved for any point of the phase diagram, i.e., also for $\Delta_{\rm eff}>0$,  by sympathetically cooling the chain \cite{Sympathetic}, corresponding to an external reservoir with $\Gamma_n>0$ in Eqs. \eqref{eq:a}-\eqref{eq:b}.

\begin{figure}
\includegraphics[width=3.4in]{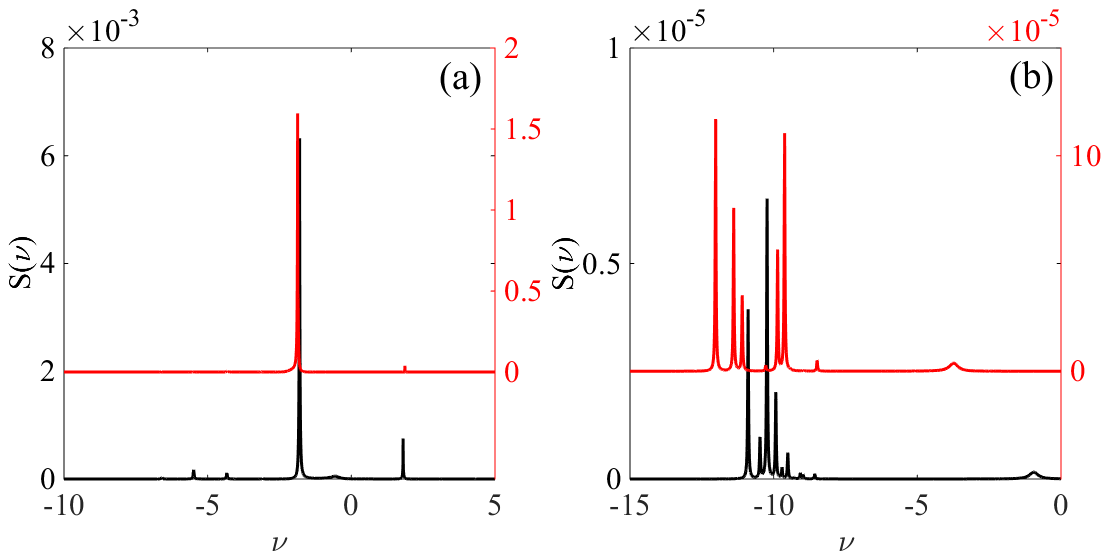}
\caption{Spectrum at the cavity output $S(\nu)$ (in arbitrary units) for $C<0$: (a) in the  sliding phase with $B_N=1-0.45$ and (b) in the  pinned phase with $B_N=1-0.05$ and for $C=-0.5$ (black line) and $C=-2$ (red line), for $11$ $^{174}$Yb$^+$ ions, $\Delta_c=0$, $\Gamma_n=0.1\kappa$, and $T=100\mu K$. The resonances correspond to vibrational eigenmodes coupling with the cavity field and change in the pinned phase as $C$ is increased. The elastic peak at $\omega_p$ (corresponding to $\nu=0$) is not reported.}
\label{Fig:3}
\end{figure}


The  spectrum of the field emitted by the cavity, $\hat a_{\rm out}$, contains information about the collective vibrational modes of the ions within the resonator \cite{Dantan,Brahms}. The output field is formally connected to the cavity field via the relation $\hat a_{\rm out}=\hat a_{\rm in}+\sqrt{2\kappa}\hat a$ \cite{Collett}, and its power spectrum is given by $S(\nu)\propto\langle \hat{\tilde{a}}_{\rm out}(\nu)^{\dagger}\hat{\tilde{a}}_{\rm out}(\nu)\rangle$, where $\hat{\tilde{a}}_{\rm out}(\nu)$ 
is the Fourier transform of $\hat a_{\rm out}(t)$ \cite{Cormick,SI}. Figure \ref{Fig:3} displays $S(\nu)$ for parameters such that the ions phases are sliding (a) 
and pinned (b). For each phase we took the same values of $B_N$ but different values of $C$. The peaks correspond to vibrational modes coupled to the cavity and it is apparent that in the pinned phase more peaks are visible. This is a result of the broken 
symmetry induced by the optical lattice potential. The effect of the cavity back-action is weak in the sliding phase 
as the only discernible change in the spectrum is its relative intensity. In the pinned phase, however, the intricacies of 
the back-action are particularly apparent. Here the spread of the cavity frequencies become more separated for 
$C=-2$ due to the softening of the cavity pinning (see Fig. \ref{fig:1}(b)), resulting in the emergence 
of three distinct frequency bands. Contrary to this, when $C=-0.5$ the ions are tightly restricted to the 
potential minima resulting in a narrow frequency band. 

Our analysis is performed for parameters which are consistent with  ongoing experiments, joining trapped ions and CQED setups \cite{Leroux,Cetina,Vuletic,Northup,Northup2} where the nonlinearity can be experimentally tuned by changing the number of atoms. This study is an example of competing long-range self-organization mechanisms which realizes a new paradigm of the Frenkel-Kontorova model. As well as displaying novel phases, such as bistability induced by the cavity mediated interactions, the inherent losses from the cavity can cool the ions in a controlled manner and allow one to monitor the phases at the cavity output, thus setting the basis for feedback mechanisms controlling the thermodynamics of friction.


\begin{acknowledgments}
We thank M. Henkel, K. Rojan, and R. Bennewitz for discussions. Financial support by the German Research 
Foundation (DFG, DACH project "Quantum crystals of matter and light"), by the French government via the 2013-2014 Chateaubriand fellowship, and by Harvard-MIT CUA, NSF Grant No. DMR-07-05472, AFOSR Quantum Simulation MURI, the ARO-MURI on Atomtronics, and the ARO MURI Quism program, is acknowledged. V.M.S. was supported by the SNSF.
\end{acknowledgments}

\section{Supplementary information}

\subsection{The model}

We treat the cavity mode in second quantization, with  $\hat n=\hat a^{\dagger}\hat a$ the photon number operator, $\hat a$ the bosonic operator destroying a quantum at energy $\hbar\omega_c$ and $\hat a^{\dagger}$ its adjoint. The resonator is driven by a classical field at frequency $\omega_p$, and we report the dynamics in the reference frame rotating with $\omega_p$.  The ions behave as classically polarizable particles, which is valid when the atomic transition which couples with the cavity mode is far detuned from the fields, so that the detuning $\Delta_0=\omega_p-\omega_{\rm el}$ is the largest frequency of the problem. Note that since $|\Delta_0|\gg |\Delta_c|$, in the parameter regimes we consider, $\Delta_0\simeq  \omega_c-\omega_{\rm el}$.
The coherent dynamics of $N$ ions and cavity field mode is governed by the Hamilton operator ($\hbar=1$)
\begin{equation}
\hat H=\sum_j\frac{\hat p_j^2}{2m}+V_{\rm ion}-\Delta_c\hat n + U_0\hat n \sum_j \cos^2k\hat x_j -{\rm i} (\eta\hat a -\eta^*\hat a^{\dagger})\,.
\end{equation}
The effect of cavity losses and damping on the motion is introduced by means of Heisenberg-Langevin (HL) equation. For the moment we assume that the only incoherent effects are losses of the cavity field at rate $\kappa$, so that the HL equation for the cavity field reads
\begin{equation}
\label{eq:a:0}
\dot {\hat a}=\frac{1}{\rm i}[\hat a,\hat H]-\kappa \hat a+\sqrt{2\kappa}\hat a_{\rm in}(t)\,.
\end{equation}
We perform the study in the semiclassical limit, assuming that the fluctuations about the mean values of field and atomic variables are sufficiently small to justify the treatment.  To this aim, we decompose the operators as a sum of mean values and fluctuations according to the prescription 
\beq
\label{eq:mean:delta}
\begin{array}{l}
\hat  a = \bar a + \delta \hat a \,,\\
\hat x_j = \bar x_j + \delta \hat x_j \,,\\ 
\hat p_j = \bar p_j + \delta \hat p_j \,,\end{array}
\eeq
where $\langle \hat a \rangle= \bar a$,   $\langle\hat x_j \rangle= \bar x_j $, and $\langle \hat p_j \rangle= \bar p_j$, while the expectation value of the fluctuations $\delta \hat a, \delta \hat x_j, \delta \hat p_j$ vanishes. \\

\begin{figure}
\includegraphics[width=3.2in]{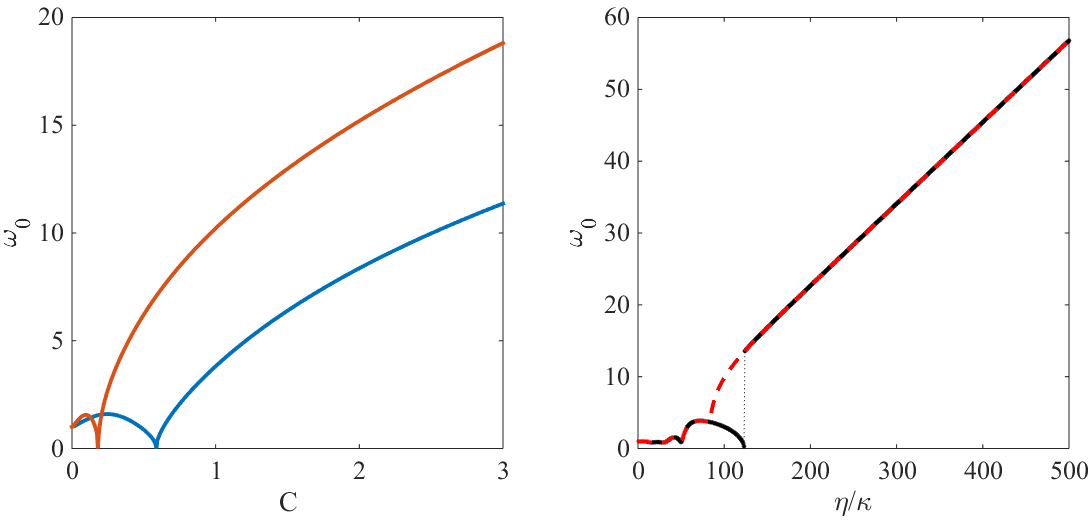}
\caption{(a) The phonon gap plotted as a function of $C$ with $\eta=100\kappa$ (blue line) and $\eta=150\kappa$ (orange line). The vanishing of the phonon gap happens at the sliding to pinned transition. (b) The phonon gap as a function of $\eta/\kappa$ for $C=2.4$ showing bistable states where a sliding (black line) and pinned (red dashed line) phase co-exist. The parameters are the same as those in Fig.2 in the main article.}
\label{phonons}
\end{figure}

{\it Mean values.} The mean values satisfy the equations of motion
\begin{align}
 & \frac{\partial~}{\partial t} \bar a = ( {\rm i} \Delta_{\rm eff} - \kappa) \bar a + \eta \,, \label{eq:a eq}\\
 & \frac{\partial~}{\partial t} \bar x_j = \frac{\bar p_j}{m} \,, \label{eq:r eq} \\
 & \frac{\partial~}{\partial t} \bar p_j = - \partial_j V_{\rm ions} - U_0 \bar n \partial_j\cos^2(k\bar x_j) \,, \label{eq:p eq}
\end{align}
with $\bar n=|\bar a|^2$ and $\partial_j=\partial/\partial x_j$ the gradient with respect to the spatial coordinates of the $j$-th particle (evaluated at the equilibrium positions $\bar x_1,\ldots, \bar x_N$), while
\beq
\label{Delta:eff}
\Delta_{\rm eff}=\Delta_c-U_0NB_N\,.
\eeq
In order to determine the classical equilibrium values we require that the quantities $\bar a$, $\bar x_j$ and $\bar p_j$ correspond to stationary solutions of the dynamical equations, namely $\partial_t \bar a=0$, $ \partial_t \bar x_j =0$, and $\partial_t \bar p_j =0$. 
From \eqref{eq:a eq} we obtain
\beq
\label{bar:a}
\bar a=\frac{\eta}{\kappa-{\rm i}\Delta_{\rm eff}}
\eeq
and with no loss of generality, we choose the phase of $\eta$ such that $\bar{a}$ is real.
Setting Eq.~(\ref{eq:r eq}) to zero gives $\bar{p}_j=0$. Substituting the value of $\bar n=|\bar a|^2$ into Eq. (\ref{eq:p eq}), one finds that it can be cast in the form
\beq 
 \frac{\partial~}{\partial t} \bar{p}_j = - \partial_j V_{\rm ions} - \partial_jV_{\rm cav}\,,
\eeq 
with 
\beq
\partial_jV_{\rm cav}=-2\frac{\eta^2}{\kappa^2} \frac{U_0\sin(k\bar x_j)\cos(k\bar x_j)}{1+(\Delta_c-U_0\sum_\ell\cos^2(k\bar x_\ell))^2/\kappa^2}
\eeq
which gives
\beq \label{eq:eff opt potential}
V_{\rm cav} = \frac{ |\eta|^2}{\kappa} \arctan \left(-\frac{\Delta_{\rm eff}}{\kappa} \right)\,.
\eeq

The equilibrium positions of the ions are then found by minimizing the total potential $V=V_{\rm ion}+V_{\rm cav}$ such that the forces due to the Coulomb repulsion, the confining potential and the cavity field are balanced. 

{\it Phonon gap.}  We determine the phonon gap by evaluating the dispersion relation of the ions vibrations in the total potential $V$ (thus discarding the fluctuations of the cavity field but taking into account its mean effect). The phonon gap is plotted in Fig.~\ref{phonons}(a) for $N=11$ ions as a function of $C$. As noted in Ref. \cite{Mata} the phonon gap does not vanish in the sliding phase as in the FK model due to the confining harmonic potential, however as the depth of the cavity potential is increased the phonon gap can fluctuate in the sliding phase. The transition from sliding to pinned can be visualized by the phonon gap vanishing at a critical cavity depth, after which the gap increases monotonically \cite{PeyrardAubry}. It should be noted that for an even number of ions a similar behaviour is observed however the phonon gap may not vanish around the transition point. This is a finite-size effect. 

The phonon gap across the bistable region is plotted in Fig.~\ref{phonons}(b) dependent on $\eta$ in units of $\kappa$. The black line shows an extended sliding phase due to the ions in the chain becoming stable as a result of the cavity back-action flattening the shape of the cavity potential. At the critical point the phonon gap increases abruptly signifying a drastic jump from the sliding to the  pinned phase. The red dashed line shows the other stable configuration for the same paramaters, however in this case the sliding phase is shorter as the ions move to the pinned phase at a lower critical value of the cavity depth.

{\it Fluctuations.} The coupled dynamics of the quantum fluctuations of field and motion are governed by the HL equations \cite{Collett}, which are found substituting the decomposition Eq.~\eqref{eq:mean:delta} into Eq.~\eqref{eq:a:0} and into the Heisenberg equations of motion for the center-of-mass variables, and using that the mean values are the stationary solutions. The equations read 
\begin{align}
\label{eq:a fluct}
\delta\dot{\hat a} &= ({\rm i} \Delta_{\rm eff} - \kappa) \delta \hat a - {\rm i}U_0 \bar a \sum_\ell (\delta \hat x_\ell \partial_\ell)\cos^2(k\bar x_\ell) + \sqrt{2\kappa}\, \hat a_{\rm in}\,, \\
\label{eq:r fluct}
\delta\dot{\hat x}_j&= \frac{\delta \hat p_j}{m}\,, \phantom{\sum_k}\\
\label{eq:p fluct}
\delta\dot{\hat p}_j &= - \sum_\ell(\delta \hat x_\ell \partial_\ell) \left( \partial_j V_{\rm ions} + U_0 \bar n\partial_j\cos^2(k\bar x_j) \right) \nonumber \\
& \quad \, - U_0\left({\bar a}^* \delta \hat a + \bar a \delta \hat a^\dagger\right) \partial_j\cos^2(k\bar x_j)  \,,
\end{align}
where the derivatives in the expressions above are evaluated at the equilibrium positions. 

For convenience we introduce the normal modes of the crystal, that characterize the dynamics of the ions when the coupling with the quantum fluctuations of the cavity field can be neglected: 
\beq
\delta \hat x_j = \sum_n M_{jn} \sqrt{\frac{\hbar}{m\omega_n}} \hat q_n \, , 
\eeq
with $M_{jn}$ the element of the orthogonal matrix relating the local coordinates $\delta \hat x_j$ with the normal-mode coordinates, 
that diagonalize Eqs. (\ref{eq:r fluct})-(\ref{eq:p fluct}) when the cavity fluctuations $\delta \hat a$ are set to zero; 
the variables $\hat q_n$ are the dimensionless position coordinates of the normal modes.  

We denote by $b_n$ and $b_n^\dagger$ the bosonic operators annihilating and creating, respectively, a phonon of the 
normal mode at frequency $\omega_n$. They are defined through the equations $\hat q_n=(\hat b_n+\hat b_n^\dagger)/\sqrt{2}$ and 
$p_n= {\rm i}(\hat b_n^\dagger-\hat b_n)/\sqrt{2}$, and the dynamical equations for the fluctuations then take the form:
\begin{align}
\label{eq:a}
\!\!\!& \delta \dot{\hat a} = ({\rm i} \Delta_{\rm eff} - \kappa) \delta  \hat a - {\rm i} \bar a \sum_n c_n ( \hat b_n +  \hat b_n^\dagger) + \sqrt{2\kappa}\,  \hat a_{\rm in}\,, \\
\label{eq:b}
\!\!\!& \dot{\hat b}_n= - ({\rm i} \omega_n + \Gamma_n)  \hat b_n - {\rm i} \bar a c_n (\delta  \hat a + \delta  \hat a^\dagger) + \sqrt{2\Gamma_n} \,  \hat b_{{\rm in}, n}\,,
\end{align}
which also includes the coupling of mode $n$ to a reservoir at rate $\Gamma_n$. The corresponding Langevin force is described by the input noise operator $\hat b_{{\rm in}, n}$, with $\langle \hat b_{{\rm in},n} \rangle = 0$ and
\beq \label{eq:b-input}
\langle \hat b_{{\rm in}, n}^\dagger(t') \, \hat b_{{\rm in}, n'} (t'') \rangle = \bar N_n \, \delta_{n n'} \, \delta(t'-t'')\,.
\eeq
with $\bar N_n = \bar N (\omega_n)$ the mean excitation number of an oscillator of frequency $\omega_n$ at the  temperature of the considered environment.
The coefficients $c_n$ in Eq. \eqref{eq:a}-\eqref{eq:b} read
\beq
c_n = \sqrt{\frac{\hbar}{2m\omega_n}} U_0\sum_j M_{jn} \partial_j \cos^2(k\bar x_j)\,, \label{eq:cavity-motion-coupling}
\eeq
where the derivatives are evaluated at the equilibrium positions $x_j$. 

\begin{figure}
\includegraphics[width=80mm]{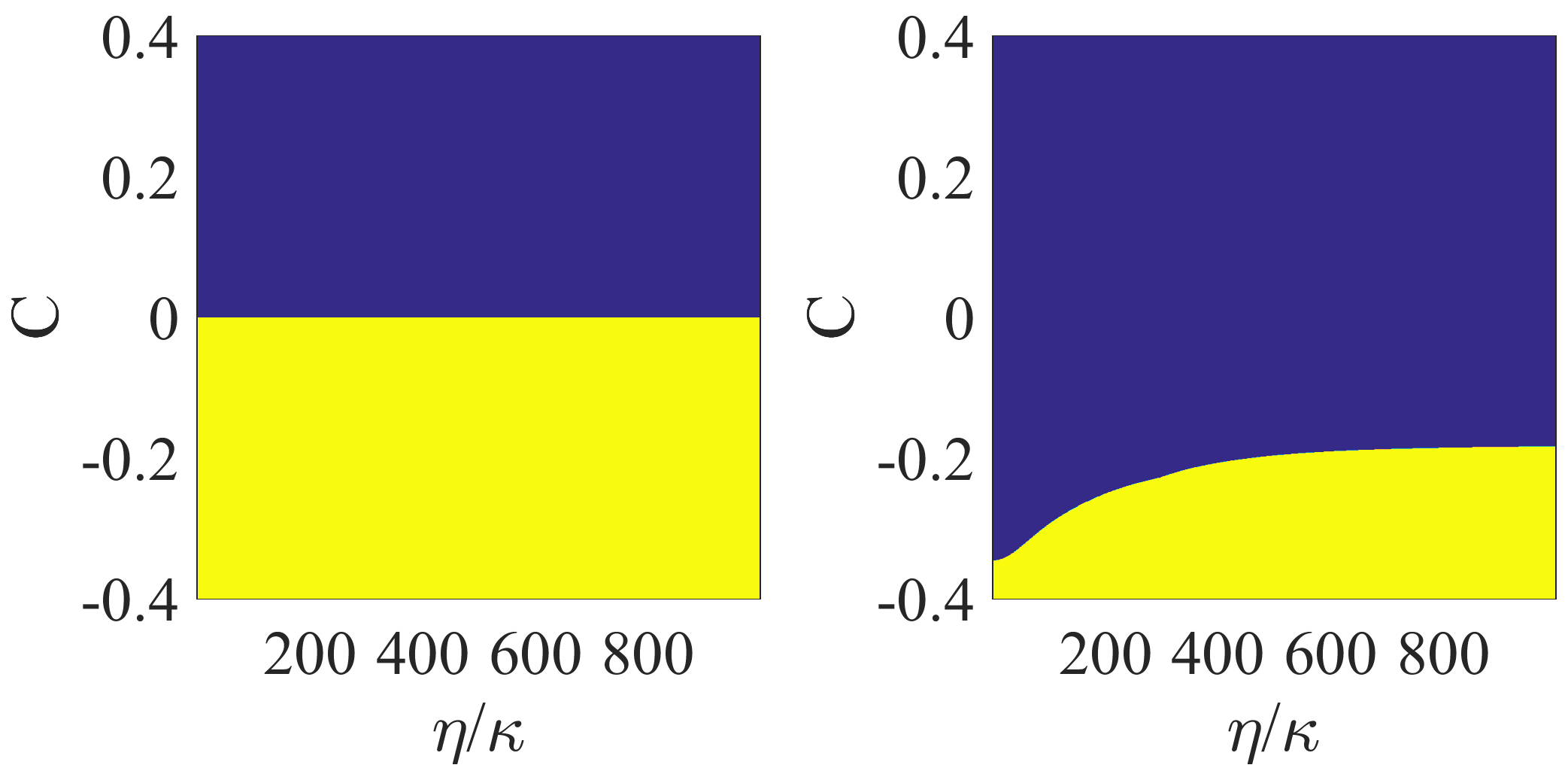}
\caption{Unstable (yellow) and stable (blue) regions of the system for (a) $\Delta_c=0$ and (b) $\Delta_c=-2\kappa$. For the system to be stable the eigenvalues of the matrix $A$ (see Eq.~$42$ in \cite{Cormick}) must have negative real parts. If this is not the case then there will be at least one mode which is heated by the cavity. }
\label{stab}
\end{figure}

\begin{figure}
\subfigure{\includegraphics[width=41mm]{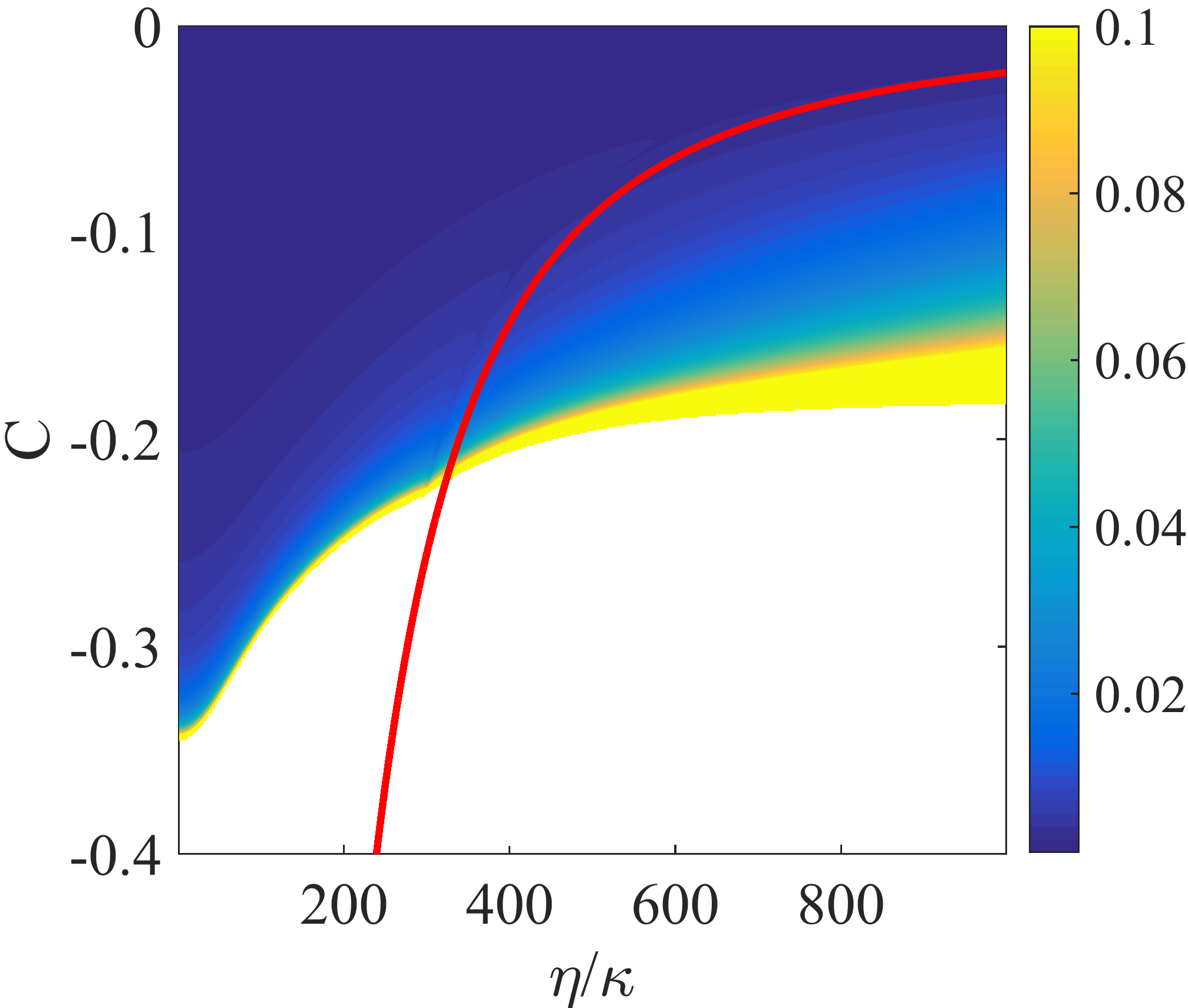}}
\subfigure{\includegraphics[width=41mm]{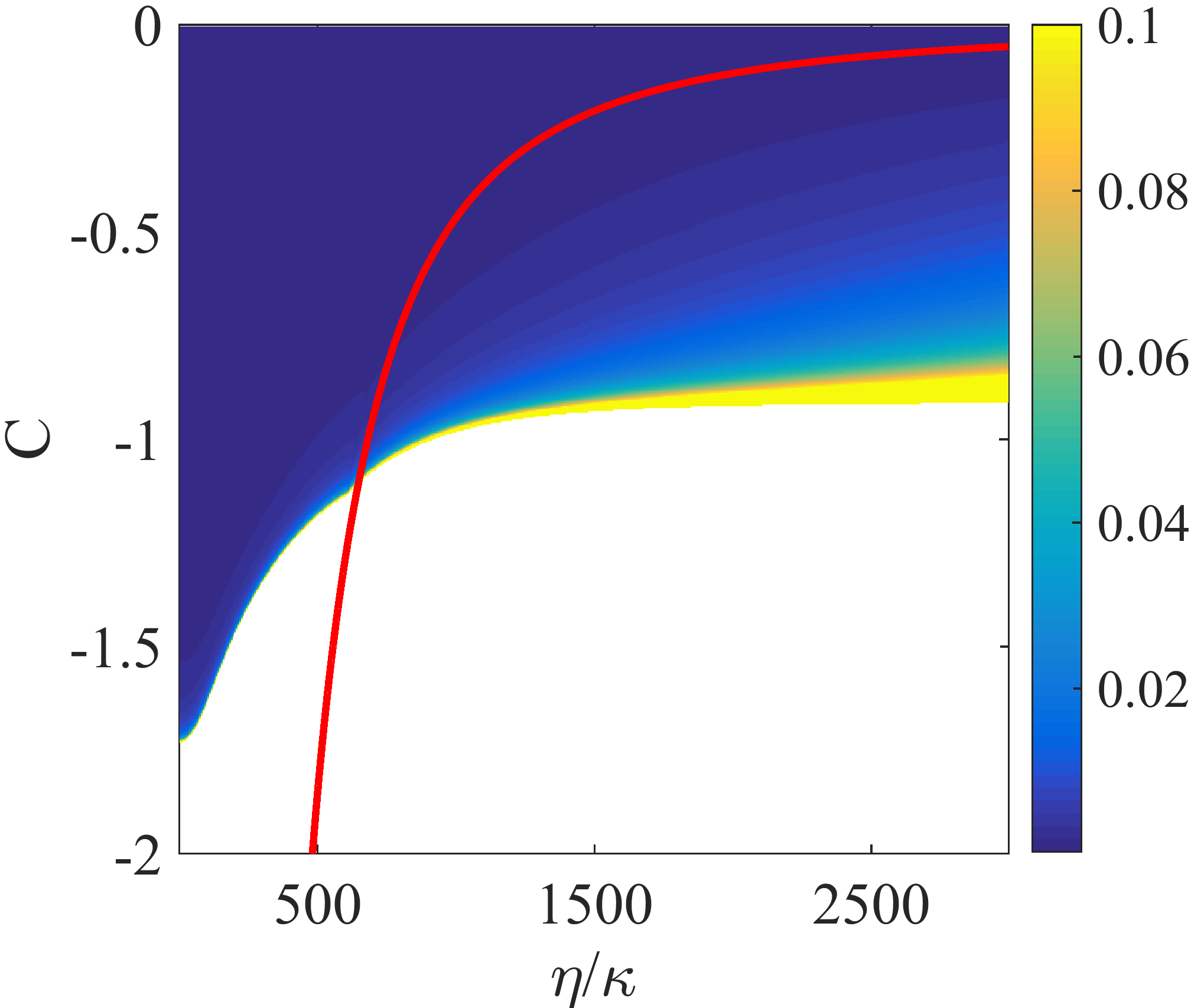}}
\caption{Average chain temperature in units of degrees Kelvin versus $C$ and $\eta$ in units of $kappa$. The chain is cooled in the colored area by the cavity, which is reached by suitably modifying the detuning, shown for (a) $\Delta_c=-2\kappa$ and (b) $\Delta_c=-10\kappa$. The solid red line is the symmetry breaking transition, and the white area is the region of the parameter space which experiences heating from the cavity.}
\label{temp}
\end{figure}

{\it Stability diagrams and cavity cooling}
The stability of the system is dependent on the equations of motion coupling the cavity and motional fluctuations which can be written in the compact form
\begin{equation}
\frac{d\overrightarrow{X}}{dt}=A\overrightarrow{X}+\overrightarrow{X}_{in}(t)
\end{equation}
where the elements of the matrix $A$ contain the arguments of Eq.~\eqref{eq:a} and Eq.~\eqref{eq:b} \cite{Cormick}. 
For the system to be stable the real parts of the eigenvalues of $A$ must be negative, otherwise at least one 
mode of the chain will be heated. It can be shown that the stability condition is $\Delta_{\rm eff}<0$. The stable regions of our parameter space are plotted in Fig.~\ref{stab} for zero and finite $\Delta_c$, with $\Gamma_n = 0 \, \forall\, n$. In the extended stable region produced by the detuning the chain is cooled by the cavity to sufficiently low temperatures (minimum of around $125\mu$K for $\Delta_c=-10\kappa$) and this is shown in Fig.\ref{temp}. The cavity cooled region may be increased by selecting a larger $\Delta_c$, and the transition may be observed as a function of $\eta$ by choosing specific $C$ thereby only experiencing a small increase in temperature. 

In order to stabilise the rest of the unstable region, all the modes can be coupled to an environment that can damp the excitations (this coupling is denoted by a finite $\Gamma_n$). To stabilize the entirety of the unstable regions shown here we must choose the coupling to be $\Gamma_n=0.1\kappa$ with the temperature of the external bath being $T_{\rm ext}=100\mu K$.

\subsection{Evaluation of the restoring force}
\begin{figure}
\includegraphics[width=3.2in]{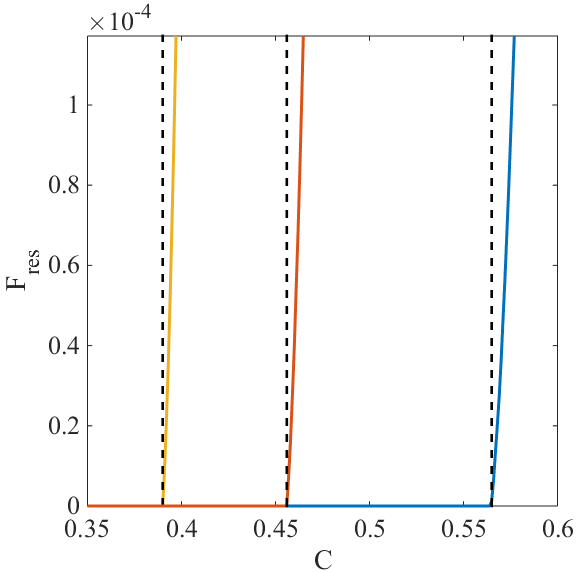}
\caption{Restoring force in units of $m\omega^2L$ versus $C$, where $L=(q^2/4\pi\epsilon_0 m \omega^2)^{1/3}$ and is of the order of the interparticle distance at the chain center $d$. Three different pump strengths are shown (from left to right): $\eta=100\kappa$ (blue), $\eta=105\kappa$ (red) and $\eta=110\kappa$ (yellow) 
(the critical value of C becomes smaller with increasing $\eta$). The symmetry breaking transition is also shown as a black dashed line for each case. The parameters are the same as those in Fig.2 in the main article.}
\label{resF}
\end{figure}

To discern the sliding and pinned phases we calculate the restoring force required to return the central ion back to the maxima of the cavity field. This is analogous to the de-pinning force of the Frenkel-Kontorova model, whereby in the sliding phase with the symmetry of the system intact the restoring force is zero. However in the pinned phase the restoring force becomes finite thereby realizing the growth of static friction in the system. A finite force is applied to all the ions equally which may be physically implemented by tilting the lattice potential \cite{Tosatti}. The sum of the total forces acting on all the ions must vanish to ensure a stable solution and are given by
\begin{equation}
F=F_{\rm ion}+F_{\rm cav}+F_{T}=0
\end{equation}
where $F_{\rm ion}=-\frac{\partial V_{\rm ion}}{\partial x}$ and $F_{\rm cav}=-\frac{\partial V_{\rm cav}}{\partial x}$ and $F_{T}$ is the finite force applied to the chain. The critical value of the restoring force is then taken as the finite force required to move the central ion back to the maximum of the cavity field, i.e. such that $\cos^2(k x_0)=1$ at which point $F_{\rm res}=F_{T}$. The restoring force is shown in Fig.~\ref{resF} as a function of $C$ for different pump strengths $\eta/\kappa$. Before the transition the restoring force is zero, however after the transition the restoring force becomes finite highlighting the broken symmetry of the system.

\subsection{Spectrum at the cavity output}

We report here the analytical results, which were obtained in Ref. \cite{Cormick}. The spectrum is calculated after evaluating the Fourier transform of the linearized HL for the fluctuations, using $\hat a=\bar a+\delta \hat a$. The quantum component of the spectrum reads
\beq
S(\nu) =\frac{\langle \delta \hat{\tilde{a}} (\nu)^\dagger \delta\hat{\tilde{a}} (\nu) \rangle}{\bar a^2} \,,
\eeq
where $\delta\hat{\tilde{a}} (\nu)$ is the Fourier transform of $\delta \hat a$, $\delta \hat{\tilde{a}} (\nu)^\dagger$ is the Hermitian conjugate of $\delta \hat{\tilde{a}} (\nu)$, and we omit the Rayleigh peak at $\nu=0$, i.e., $\omega=\omega_p$, which corresponds to the classical part. After some algebra, one finds the expressions
\begin{eqnarray}
S(\nu)&=&S_0(\nu)\Bigg(\frac{4\kappa|\theta(\nu)|^2 \bar{a}^2}{\kappa^2+(\nu-\Delta_{\rm eff})^2}\nonumber \\
&&+\sum_n c_n^2\Gamma_n^2\left(\frac{\bar{N}_n}{\Gamma_n^2+(\omega_n-\nu)^2}+\frac{\bar{N}_n+1}{\Gamma_n^2+(\omega_n+\nu)^2}\right)\Bigg)\nonumber\\
\end{eqnarray}
where the first term is due to coupling of the quantum vacuum with the crystal vibrations with $\theta(\nu)=\sum_n \frac{c_n^2 \omega_n}{\omega_n^2+(\gamma_n-i\nu)^2}$
and the second term is due to thermal noise coupling to the modes. The prefactor is given by
\begin{equation}
S_0(\nu)=\frac{2}{\kappa^2+(\nu+\Delta_{\rm eff})^2}\left\vert 1+\frac{4 \theta(\nu)\Delta_{\rm eff}\bar{a}^2}{(\kappa-i\nu)^2+\Delta_{\rm eff}^2}\right\vert^{-2}.
\end{equation}
These are the formula we employed to evaluate Fig. 3.


\end{document}